\documentstyle[preprint,aps]{revtex}
\begin{document}
\draft
\preprint{MKPH-T-94-13}
\title{Two-body Mechanisms in Pion Photoproduction on the Trinucleon}
\author{S. S. Kamalov\cite{Sabit} and L. Tiator}
\address{Institut f\"ur Kernphysik, Universit\"at Mainz, 55099 Mainz, Germany}
\author{C. Bennhold}
\address{Center of Nuclear Studies, Department of Physics, The George
Washington University, Washington, D.C., 20052}
\date{\today}
\maketitle

\begin{abstract}
A breakdown of the Impulse Approximation is studied in pion photoproduction
on $^3$He at high momentum transfers. The usual DWIA formalism with
Faddeev wave functions which works well for small momentum transfers
deviates from experimental measurements by up to two orders of magnitude for
$Q^2>6 fm^{-2}$. It is found that the explicit inclusion
of two-body mechanisms, where the photon is absorbed on one nucleon and the
pion is emitted from another nucleon
can restore agreement with the data.
\end{abstract}

\narrowtext

Reactions on the trinucleon are an ideal testing ground to search of effects
that go beyond the Impulse Approximation since realistic correlated
three-body wave
functions are available that are reliable even at high nuclear momentum
transfers.
Previous theoretical investigations~\cite{KTB91,KTB93,Che94}
of pion scattering
and pion photoproduction on the trinucleon systems were mostly based on the
multiple-scattering approach carried out in momentum space. Using the
impulse approximation (IA) and realistic nuclear wave functions the
nonlocalities of the pion-nuclear interaction and exact treatment  of
Fermi motion have been taken into account.
Within a coupled-channels framework it has become possible
to consistently describe $\pi^+$ and
coherent $\pi^0$ photoproduction as well as elastic and charge-exchange
pion scattering on $^3$He and $^3$H and obtain a good description of
experimental measurements at momentum transfers of $Q^2<6 fm^{-2}$.
Thus, the calculations have reached a point where the conventional
one-body aspects are treated on a rather accurate level.

However, at higher momentum transfers large discrepancies appear between
measurements and theoretical calculations~\cite{KTB91}. In the region of
$Q^2>8 fm^{-2}$, calculations dramatically fail to explain existing
$^3He(\gamma,\pi^+)^3H$ data, underestimating them by up to two orders of
magnitude.
Similar discrepancies were found in coherent $\pi^0$ photoproduction
on $^3He$ and $^3H$\cite{Bel84},
in elastic pion scattering at
backward angles~\cite{KTB93,scott94}, as well as in
pion single charge exchange and pion induced eta production on
$^3$He at higher energies~\cite{KTB93,Peng87}.

In general, it has been known for a long time that imposing gauge
invariance in electromagnetic reactions with the nucleus
will generate two-body meson exchange currents via,
e.g., the Siegert theorem.  Since the bound nucleons are off-shell, the
one-body currents are no longer conserved and, therefore, meson exchange
currents are required in order to fulfill gauge invariance.
Very recently, this method was demonstrated on the specific reaction of
pion photoproduction on finite nuclei\cite{Fr94}.
Using general requirements such as current conservation in the nuclear
electromagnetic vertex ref. \cite{Fr94} obtained general expressions for
two-body correction terms similar to meson exchange currents in electron
scattering.  However, it is also well-known that this procedure generates
only the convection part of the two-body currents since the magnetic parts
are transverse by definition and, therefore, fulfil gauge invariance
separately.

The goal of this work is to study genuine two-body mechanisms in pion
photoproduction on $^3$He which do not appear in a standard distorted wave
impulse approximation framework (DWIA).
Our main requirement is to derive such two-body operators in a way that
is consistent with the usual one-body operator describing the process on the
single nucleon.
This consistency requirement can be satisfied straightforwardly by
starting from an effective Lagrangian for the
pion-nuclear production process
and to introduce the electromagnetic field by
minimal substitution. This method not only guarantees
gauge invariance but also allows the explicit calculation of contributions
from the magnetic moments.

Using the impulse approximation for the nuclear pion emission amplitude one
can write
\begin{equation}
T_{\pi}=\frac{f}{m_\pi}\,\int d\vec{r}_{\pi}
\phi_{\alpha}^{\dagger} (\vec{r}_{\pi}) < f \mid \sum^A_{j=1}
\vec{\sigma}\cdot\stackrel{\leftarrow}{\nabla}_{\pi}\,\tau_{\alpha}(j)\,\delta
(\vec{r}_{\pi}-\vec{r}_j)
\mid i > \,,
\end{equation}
where $f/m_\pi=g/2M$ with $g^2/4\pi=14$, $M$ denotes the nucleon mass,
$\vec{\tau}$ is the nucleon isospin operator and
$\phi_{\alpha} (\vec r)=\varphi_{\alpha}e^{i\vec{q}\cdot\vec r}$
is the pion wave function with the three isospin components $\alpha=1,2,3$.

The main part of the charged pion photoproduction amplitude
- seagull and pion exchange terms - can be obtained by minimal substitution
$\vec{\nabla}_{\pi}\rightarrow\vec{\nabla}_{\pi}-ie\vec{A}$
(where $\vec{A}=\vec{\epsilon}e^{i\vec{k}\cdot\vec{r}}$ is the
electromagnetic vector potential with polarization vector $\vec{\epsilon}$
and photon momentum $\vec{k}$) in the
$\vec{\sigma}\cdot\stackrel{\leftarrow}{\nabla}_{\pi}$-operator and in the
pion wave function. Since the corresponding procedure is well known and
straightforward, we present here only
the treatment of the more complicated nucleon pole (dispersive or
two-step) terms.  In our approach, we construct this part
of the pion photoproduction operator by introducing
the electromagnetic field in the nuclear wave function which satisfies the
Schr\"odinger equation. Then the initial nuclear wave function can be written
as
\begin{equation}
\mid i > \rightarrow \mid i > + \frac{1}{{\cal E}_i-\hat{H}_0}\hat{H}_{em}
\mid i>\,,\quad \hat{H}_{em} = \frac{ie}{2M}\sum^A_{n=1}
(\vec{\nabla}_{n}\cdot\vec{A}+\vec{A}\cdot\vec{\nabla}_{n})\,,
\end{equation}
where $\hat{H}_0=\sum^A_{n=1}{\hat{\vec{p}_n}}^2/2M$ is the Hamiltonian
for noninteracting nucleons.
Note that at this step in the derivation of Eq. (2) we
retained only the electromagnetic convection current which is crucial
for gauge invariance and
neglected minimal substitution in the nucleon-nucleon interaction.

In order to get a standard one-body operator used in the IA\cite{Bl77}
we will apply the closure approximation in the evaluation of the nuclear
propagator
$1/({\cal E}_i-\hat{H}_0)$ with the mean nuclear excitation energy
${\bar E}$:
\begin{equation}
\frac{1}{{\cal E}_i-\hat{H}_0}\rightarrow\frac{1}{{\cal E}_i-{\bar E}} =
\frac{2M}{2ME_{\gamma}-\vec{k}^2}\,.
\end{equation}
Thus, in the expansion of the
nuclear propagator only the leading term was retained and the
difference between the nuclear ground and excited states energies was
neglected.

Substituting Eqs. (2,3) in Eq. (1) yields
\begin{equation}
T_{\pi\gamma}^{(s)}=\frac{-i e f}{m_\pi}\varphi_{\alpha}^{\dagger}
< f \mid \sum^A_{j=1}\hat{\tau}_{\alpha}(j)\vec{\sigma}_{j}
\cdot\vec{q}\,e^{-i\vec{q}\cdot\vec{r}_j}\sum^A_{n=1}\hat{e}(n)\,
e^{i\vec{k}\cdot\vec{r}_{n}}\frac{(2\hat{\vec p}_n+
\vec{k})\cdot\vec\epsilon}{2ME_{\gamma}-\vec{k}^2} \mid i > \,,
\end{equation}
where the operator $\hat{\vec p}_n=-i\vec\nabla_n$ acts on the initial
nuclear state $\mid i>$ and is associated with the initial nucleon momentum
$\vec{p}_i$. The isospin operator for the nucleon charge is
$\hat{e}=(1+\tau_3)/2$.

The expression for the dispersive term in Eq. (4) contains
matrix elements of one-body as well as two-body operators. The former
corresponds to the case $n=j$ and is shown in Fig. 1a.
This term is identical to the $s$-channel amplitude in pion photoproduction
on a single nucleon\cite{Bl77}. The two-body part of Eq. (4)
corresponds to the case $n \neq j$ and is shown in Fig. 1c.
This corresponds to a new class of diagrams that do not
appear in the IA.

Applying the same procedure described above for the final nuclear state
$\mid f>$ we also obtain the pion photoproduction amplitude in the
$u$-channel,
\begin{equation}
T_{\pi\gamma}^{(u)}=\frac{i e f}{m_\pi}\varphi_{\alpha}^{\dagger}
< f \mid \sum^A_{n=1}\hat{e}(n)\,\frac{(2\hat{\stackrel{\leftarrow}{p}}_n
-\vec{k})\cdot\vec\epsilon}{2ME_{\pi}+\vec{q}^2}e^{i\vec{k}\cdot\vec{r}_{n}}
\sum^A_{j=1}\hat{\tau}_{\alpha}(j)\vec{\sigma}_{j}\cdot\vec{q}\,
e^{-i\vec{q}\cdot\vec{r}_j} \mid i > \,,
\end{equation}
where the operator $\hat{\stackrel{\leftarrow}{p}}_n =
i\stackrel{\leftarrow}{\nabla}_n$ acts on the final
nuclear state $\mid f>$ and is associated with the final nucleon momentum
$\vec{p}_f$. Again, the one-body part of this amplitude (case $n=j$) describes
the elementary process in the $u$-channel (Fig. 1b),
while the two-body part  (case $n \neq j$) is shown in Fig. 1d.

Finally, choosing the Coulomb gauge, $\vec{k}\cdot\vec{\epsilon}=0$, we
obtain the standard expression for the one-body part of the  dispersive
amplitude which is traditionally used in the IA.
Furthermore, we generate a novel two-body mechanism, that is
given by
\begin{eqnarray}
T_{\pi\gamma}^{conv.}(2)=T_{\pi\gamma}^{(s)}(2)+T_{\pi\gamma}^{(u)}(2)& =
&\frac{-i e f}{m_\pi}\varphi_{\alpha}^{\dagger}
< f \mid \sum^A_{j \neq n}\hat{e}(n)\hat{\tau}_{\alpha}(j)\,\vec{\sigma}_{j}
\cdot\vec{q}\,e^{i\vec{k}\cdot\vec{r}_{n}-i\vec{q}\cdot\vec{r}_j}
\,\nonumber\\ & & \times
(\frac{2\hat{\vec p}_n\cdot\vec{\epsilon}}{2ME_{\gamma}-\vec{k}^2}-
\frac{2\hat{\stackrel{\leftarrow}{p}}_n\cdot\vec{\epsilon}}
{2ME_{\pi}+\vec{q}^2})
\mid i > \,.
\end{eqnarray}

The contributions from the
magnetic interaction due to the magnetic moments of the
proton, $\mu_p=2.79$, and the neutron, $\mu_n=-1.91$, can be obtained
analogously. The corresponding Hamiltonian is
\begin{equation}
\hat{H}_{magn.}=\frac{e}{2M}\,\sum^A_{n=1}\,\hat{\mu}(n)\,
\vec{\sigma}_n\cdot{\vec B}(\vec{r})\,,
\end{equation}
where
${\vec B}(\vec{r})=\vec\nabla\times{\vec A}=
i[{\vec k}\times\vec{\epsilon}]\,e^{i\vec{k}\cdot\vec{r}}$.
In analogy to the convection part  considered above the two-body mechanism
due to the nucleon magnetic moments reads
\begin{eqnarray}
T_{\pi\gamma}^{(magn.)}(2) & = & \frac{e f}{m_\pi}\varphi_{\alpha}^{\dagger}
< f \mid \sum^A_{j \neq n}\hat{\tau}_{\alpha}(j)\hat{\mu}(n)
\,\vec{\sigma}_{j}\cdot\vec{q}\,\vec{\sigma}_n\cdot[{\vec k}
\times\vec\epsilon]
\,\nonumber\\ & & \times
(\frac{1}{2ME_{\gamma}-\vec{k}^2}-\frac{1}{2ME_{\pi}+\vec{q}^2})
e^{i\vec{k}\cdot\vec{r}_{n}-i\vec{q}\cdot\vec{r}_j}\mid i > \,,
\end{eqnarray}
where the magnetic isospin operator is defined as
$\hat\mu=\mu_p(1+\tau_3)/2+\mu_n(1-\tau_3)/2$.

Even at large momentum transfer the energy difference
$E_{\gamma} - E_{\pi}$ is only a few MeV, therefore, to leading order the
$s$- and $u$-channel propagators cancel each other.
This has also been found by Levchuk and Shebeko\cite{Lev89} in a nuclear
pion photoproduction study that investigated effects
beyond the impulse approximation.
However, as we demonstrate below, it is exactly the difference
between the $s$- and $u$-channel contribution of order
$(\vec{k}^2+\vec{q}^2)/(2M E_\gamma)$ that is very important in the high
momentum transfer region.

We begin our discussion by considering pion rescattering effects.
Single rescattering, shown in Fig. 1e,f, is known to be incomplete
especially in the $\Delta$ resonance region, where the pion-nucleon
interaction is very strong.
Within a multiple scattering framework\cite{KMT55} we have studied
pion scattering and photoproduction
on $^3$He in detail in Refs.\cite{KTB91,KTB93}.
In Fig. 2 we show results of our previous work, comparing a PWIA
calculation without any
pion rescattering with a DWIA computation with full
pion-nucleus final state interaction
including single charge exchange. The latter one describes the data well
up to $Q^2\approx 6 fm^{-2}$. Since the standard multiple
scattering framework contains contributions from the trinucleon ground
state only, we have estimated the additional contributions
from the coupling to the break-up channels using closure approximation.
However, comparing the dashed and dash-dotted curves in Fig. 2, it is
clear that the disagreement
at high momentum transfer can not be improved significantly by
contributions from pion rescattering alone.

This situation changes dramatically, once the novel two-body mechanisms,
shown in Fig. 1c and d, are taken into account.  As shown in Fig. 2,
including these two-body
amplitudes of Eqs. (6) and (8) within the multiple scattering framework
raises the cross section by up to two orders of magnitudes, thus,
these two-body mechanisms in fact become
dominant for $Q^2>7 fm^{-2}$.
This effect removes most of the discrepancy between theory
and experiment. Our analysis indicates that this large enhancement comes mainly
from the
isovector magnetic interaction of the two-body operator defined by Eq. (8).
As mentioned above, even as the leading
terms of the $s$- and $u$-channel propagators cancel each other,
the next higher-order term, proportional to
 $(\vec{k}^2+\vec{q}^2)/(2M E_\gamma)$,
becomes significant in the high momentum transfer region.
  Note, that this cancellation of the leading two-body terms
was also found by Jennings\cite{Jenn88} in a study
of pion-deuteron scattering.
On the other hand, the two-body contribution
arising from the convection part of the electromagnetic current  (see Eq. (6))
is significantly smaller
because it does not have the enhancing factor of the isovector moment and,
moreover, is of nonlocal nature.

In Fig. 3 we illustrate the importance of the two-body mechanisms for the
differential cross sections at backward angles. Starting at a photon
energy of $E_{\gamma}=300$ MeV, the contributions of diagrams Fig. 1c,d
become visible. At $E_{\gamma}=500$ MeV and $\theta_{\pi}>150^0$,
corresponding to $Q^2>16 fm^{-2}$, the two-body mechanisms increase the
differential cross section by up to two orders of magnitude.

Finally, we demonstrate in a simple qualitative approach how momentum
sharing among the nucleons inside the nucleus enhances the cross section
in the high momentum transfer region.
In the simple harmonic oscillator model the contribution from the one-body
operator is determined by the well-known gauss form factor
\begin{eqnarray}
F^{(1)}(\vec{Q})=\int d{\vec p}\,d\vec{\cal P}\Psi_S^+(\vec{p},
\vec{\cal P}-{\textstyle \frac{2}{3}}\vec{Q})\,\Psi_S(\vec{p},\vec{\cal P})
=\,e^{-b^2Q^2/6}\,,
\end{eqnarray}
where $\vec{Q}=\vec{k}-\vec{q}$ is the nuclear momentum transfer which in
the IA has
to be absorbed by a single nucleon and
$\Psi_S$ is the $S$-shell wave function defined as
$\Psi_S(\vec{p},\vec{\cal P})=N exp(-b^2({p}^2+\frac{3}{4}
{\cal P}^2))$.
For the  matrix element that contains the two-body operator we arrive at
the expression
\begin{eqnarray}
F^{(2)}(\vec{\cal K},\vec{Q})=\int d{\vec p}\,d\vec{\cal P}
\Psi_S^+(\vec{p}+{\textstyle \frac{1}{2}}\vec{\cal K},\vec{\cal P}+
{\textstyle \frac{1}{3}}\vec{Q})\,\Psi_S(\vec{p},\vec{\cal P})
=\,e^{-b^2Q^2/24}\,e^{-b^2{\cal K}^2/8}
\end{eqnarray}
with $\vec{\cal K}=\vec{k}+\vec{q}$. Therefore, sharing of
momentum transfer among two nucleons leads to a much smaller
exponential argument compared to the one-body case.
This becomes obvious at high energies and backward angles where the
photon and pion momenta have similar magnitude but opposite sign,
leading to $\vec{\cal K}\approx 0$.
For example, in the case of pion photoproduction at
$E_{\gamma}$=400 MeV, $\theta_{\pi}=137$ and 180 we have
$F^{(2)}/F^{(1)} \approx340$ and 3000 respectively.


In conclusion, we have studied the dramatic disagreement between high-Q
($\gamma, \pi^+$) data on $^3He$ and DWIA calculations that underpredict
these data by up to two orders of magnitude.  We have resolved this
long-standing discrepancy by including explicit two-body mechanisms that
go beyond the Impulse Approximation.  These two-body terms, where the
photon is absorbed on one nucleon and the pion is emitted from another,
allow momentum transfer sharing between the two nucleons and dominate
the cross section for $Q^2 > 7 fm^{-2}$.  The most important
contribution of these two-body currents are found to be due to the
magnetic interaction.  Pion rescattering, which is important in the
kinematic region of $Q^2 < 6 fm^{-2}$, cannot account for the observed
discrepancy, even if the coupling to the break-up channels is included
in the intermediate states.

{}From our derivation it is clear that these novel two-body mechanisms are not
a special feature of the $^3He(\gamma,\pi^+)^3H$ process
but should play an important
role in all scattering and production processes at high momentum
transfer, as long as the nucleus remains in its ground state. Such
processes are coherent $\pi^0$ photoproduction on the deuteron or $^4He$,
$(\gamma,\eta)$
on light systems, nuclear Compton scattering and, furthermore,
meson elastic scattering and single charge exchange processes.
The new high energy, 100$\%$ duty factor electron
accelerators MAMI at Mainz, ELSA at Bonn and CEBAF are the ideal tools
to investigate this field of two-body effects and nucleon-nucleon correlations.

This work was supported by the Deutsche Forschungsgemeinschaft (SFB201),
the U.S. DOE grant DE-FG05-86-ER40270 and the Heisenberg-Landau program.


\begin{figure}
\caption{
Diagrams for the dispersive and pion rescattering terms
in nuclear pion photoproduction. }
\end{figure}

\begin{figure}
\caption{
Differential cross section at $\Theta_{c.m.}=137^0$ as a function of
nuclear momentum transfer $Q^2$. The dotted (dashed) curves show the PWIA
(DWIA) results obtained with Faddeev wave functions\protect\cite{Bra75}.
The dash-dotted curve includes the corrections due to the coupling with the
break-up channels
and the full line shows our complete calculation with the additional
novel two-body contribution of Fig. 1c,d. The experimental data are
from Ref.\protect\cite{Bach73} }
\end{figure}

\begin{figure}
\caption{
Pion angular distribution at $E_{\gamma}=300$, 400 and 500 MeV.
The notations of the curves are the same as in Fig. 2.
The experimental data are from Refs.\protect\cite{Bach73,Hose88}. }
\end{figure}

\end{document}